\documentclass[12pt]{article}

\usepackage{scicite}

\usepackage{times}

\usepackage{amsmath}
\usepackage{amsfonts}
\usepackage{amssymb}
\usepackage{graphicx}
\usepackage{float}

\usepackage{hyperref}
\usepackage{multirow}
\usepackage{rotating}

\usepackage{lineno}

\renewcommand{\baselinestretch}{1.5}

\newenvironment{sciabstract}{%
\begin{quote} \bf}
{\end{quote}}

\title{Differences in collaboration structures and impact among prominent researchers in Europe and North America}

\author{Lluís Danús$^{1,\ast}$, Carles Muntaner$^{2}$, Alexander Krauss$^{3,4}$,\\
Marta Sales-Pardo$^{1,\ast}$, and Roger Guimerà$^{5,1,\ast}$\\
\normalsize{$^1$Department of Chemical Engineering, Universitat Rovira i Virgili, Tarragona, Catalonia}\\
\normalsize{$^2$Dalla Lana School of Public Health, University of Toronto, Toronto, Canada}\\
\normalsize{$^3$London School of Economics, London, UK}\\
\normalsize{$^4$Universitat de Barcelona, Barcelona, Catalonia}\\
\normalsize{$^5$ICREA, Barcelona, Catalonia}\\
\\
\normalsize{$^\ast$Corresponding author(s). E-mail(s):}\\
\normalsize{lluis.danus@urv.cat, marta.sales@urv.cat, roger.guimera@urvcat} 
}

\date{}

\begin{document} 


\baselineskip24pt


\maketitle


\begin{sciabstract}
Scientists collaborate through intricate networks, which impact the quality and scope of their research. At the same time, funding and institutional arrangements, as well as scientific and political cultures, affect the structure of collaboration networks. Since such arrangements and cultures differ across regions in the world in systematic ways, we surmise that collaboration networks and impact should also differ systematically across regions. To test this, we compare the structure of collaboration networks among prominent researchers in North America and Europe. We find that prominent researchers in Europe establish denser collaboration networks, whereas those in North-America establish more decentralized networks. We also find that the impact of the publications of prominent researchers in North America is significantly higher than for those in Europe, both when they collaborate with other prominent researchers and when they do not. Although Europeans collaborate with other prominent researchers more often, which increases their impact, we also find that repeated collaboration among prominent researchers decreases the synergistic effect of collaborating.
\end{sciabstract}


\section*{Introduction}

Science is a social endeavor that progresses through the concerted effort of many individuals, who exchange ideas and interact through intricate collaboration networks  \cite{price63,merton73,newman01b,guimera05c}. 
Due to the increasing complexity involved in the most pressing problems in science and society, and the advantage of diverse groups at solving complex tasks \cite{Almaatouqe21, carletti20}, the role of these collaboration networks is becoming more and more important to achieve scientific excellence and advance research fields \cite{wuchty07,jones08}. Additionally, the structure of collaboration networks affects the quality and scope of research outcomes in different ways, some of which have been well described. In particular, networks with more recurrent collaborations have been linked to research with lower impact \cite{guimera05c, chan16}. Also, centralized scientific communities (those with a highly connected cluster in which the same group of scientists repeatedly co-author articles) are more likely to propagate non-replicable claims, and vice versa for decentralized communities with less overlap in co-authorship and more diverse methods \cite{danchev19}. The structure of the collaboration network also has an impact on the career of researchers. For example, network structure is predictive of who produces groundbreaking ideas, and who wins scientific prizes and awards \cite{ma18}.

At the same time, there is mounting evidence that different research environments, such as different funding and institutional arrangements or different scientific and political cultures, leave measurable fingerprints in collaboration networks \cite{srivastava11,uzzi13,bromham16,wu19}. For example, we know that resource-intensive fields (such as astrophysics or high energy physics) typically have collaborations involving a large number of researchers (more than 100) and, therefore, denser collaboration networks \cite{guimera05c, wagner17}. Resource demands also result in gender imbalance: women tend to be excluded from resource-intensive fields that require large collaborations (for example, genomics versus plant sciences in biology) and therefore end up working in smaller teams \cite{zeng16}.
Since funding and institutional arrangements and scientific and political cultures differ across regions in the world in systematic ways, we surmise that collaboration networks should also differ systematically across regions, independently of other factors such as research field. Additionally, because of the effect of collaboration network structure on research outcomes, we expect to observe systematic differences in the impact of research produced in North America and Europe. Such differences have indeed been observed \cite{king04,lepori19}; we explore whether they are affected by collaboration-related factors.

We address the lack of comparative studies on collaboration networks across regions by collecting data on field-specific collaboration networks for eight different fields and classifying prominent researchers
%
%
based on their institutional affiliation in one of these two regions. We do not observe systematic differences between prominent researchers in Europe and North America if we look exclusively at  the publication output and the overall number of collaborators. However, consistent with our hypothesis, we find that collaboration networks in North America and Europe do have distinctive features that are robust across fields. Specifically, we find that prominent researchers in Europe build denser collaboration networks with each other, and those in North America build more decentralized networks, with researchers in these two regions thus fulfilling different structural roles. 
Also consistent with our expectations, we observe differences in the impact of the publications of each community, and find that these differences are collaboration-dependent. In particular, the impact (normalized by publication year) of the publications of prominent researchers in North America is significantly higher (12\% overall) than for those in Europe, when they do not collaborate with other prominent researchers in the field. When prominent researchers collaborate with each other, which Europeans do more frequently, normalized impact tends to increase, overall, by 10\% for Europeans and 21\% for North Americans, broadening the impact difference to 22\% overall.



\section*{Results}

\subsection*{Collaboration networks between prominent researchers in different fields}

We start by collecting data on the scientific collaboration networks between roughly 100 prominent researchers in eight different scientific fields: genetics, development economics, cognitive psychology, philosophy of science, network science, metabolomics, network ecology, and social inequalities in health. We focus on prominent researchers for two main reasons. First, elite researchers are responsible for much of the impact and research focus in any field \cite{merton68,petersen11,barabasi12,li20,ioannidis19}. Second, since they also receive a disproportionate share of the funding in their field,  they are more likely to be  sensitive to institutional arrangements, scientific cultures and funding strategies.

We choose these eight fields because they provide a broad scope of fields spanning across the natural and social sciences and they are sufficiently small and well-defined for prominent researchers to collaborate with one another, while being sufficiently established to have a consistent track record of collaborations between prominent researchers, and of the impact of these collaborations. Moreover, these fields are diverse in terms of the topics covered, their scientific cultures, and how established they are. In particular, the first four fields have longer traditions, whereas the latter are relatively young and have evolved for shorter times. Finally, three of these fields have a majority of prominent researchers based in Europe whereas five have a majority based in North America, with overall 40\% of researchers based in Europe and 60\% in North America (Supplementary Fig.~S10).

For the field of inequalities in health, we used the list of prominent scientists assembled by another research group \cite{bouchard15}. For the other seven fields, we compiled data for the top 100 researchers using different criteria depending on the nature of the field (for example, some fields have well-defined conferences that we used, whereas others do not; see Methods). We identified the 100 with the highest H-index \cite{hirsch05}, according to the Scopus database. We excluded a few researchers  who were not based in Europe or North America and a few who did not have any collaboration with the others (these were typically prominent researchers in other fields with only a few publications in the fields we study). See Materials and Methods section in the appendix for further details. The results we report in what follows are consistent across fields, which provides assurance that our findings are robust and not dependent on the methods used to compile the networks.

Collaboration patterns, and their outcomes in terms of publications, do not appear, at first glance, to be vastly different for prominent researchers in North America and Europe (Fig.~1). In both cases, we observe large variability in the total number of collaborators 
and the total number of publications of prominent researchers. 
Because of this variability, in what follows we consider the logarithm of the number of collaborations, the number of collaborators and the impact of publications \cite{stringer08}.
As expected, we observe that the number of collaborators grows with the number of publications; but we observe no consistent significant differences between Europe and North-America (Fig.~1A-H) (except in the case of network ecology, $p=0.02$, and genetics, $p=0.01$).
%
%
By analyzing all papers of the prominent researchers, we also find no evidence that researchers in either of the two geographic regions engage in systematically larger collaborations, as measured by the number of authors per article (Supplementary Fig.~S1A-H).
%

\subsection*{Roles of prominent researchers in Europe and North America}

Despite the lack of systematic differences among geographical groups of prominent researchers in terms of the total number of publications and collaborations, a more nuanced analysis of the structure of the collaboration network between prominent researchers (that is, excluding their collaborators who are not prominent) reveals systematic and consistent differences between North America and Europe. We start by modeling the network using a hierarchical \cite{peixoto14} stochastic block model (hSBM) \cite{white76,nowicki01,guimera09} (Fig. 2A and Supplementary Figs.~S2-S9A). This approach partitions researchers into groups according to their collaboration patterns (Methods). Researchers in the same group occupy a similar position in the network and thus play a similar role \cite{white76}. Unlike other methods to identify groups, roles and/or positions in networks, our approach (Bayesian maximum a posterior, or, equivalently, minimum description length; see Methods) guarantees that the partition of the network into groups is the most parsimonious.

The groups we obtain are markedly polarized in their composition (Fig. 2B,C), with some groups containing mostly researchers in Europe and others containing mostly researchers in North America, meaning that researchers with the same structural role are typically based in the same continent. We quantify this polarization by counting the number of same-continent researchers in each group, and comparing those numbers to the null expectation obtained by resampling researchers' institutional affiliations (Methods). We find that group polarization is highly significant in all fields except philosophy of science, where the scarcity of collaborations leads to non-significant results (Fig. 2C and Supplementary Figs.~S2-S9C). This indicates that prominent researchers in North America and Europe fulfill distinct structural roles in collaboration networks between prominent researchers.

Group polarization could be naively attributed to geographic proximity, that is, to the tendency of researchers based in the same continent to collaborate;  indeed, this would lead to polarized groups. However, deeper analysis of the collaboration networks and the corresponding block models (Fig. 2A,B and Supplementary Figs.~S2-S9) reveals that this is not the only factor at play. Rather, we observe genuinely different collaboration patterns across continents. Groups with more Europe-based researchers tend to have more within-group and between-group collaborations, whereas groups with more researchers in North America tend to have fewer collaborations altogether. In the following, we quantify these differences directly in the collaboration networks between prominent researchers.

First, we measure the total number of collaborations between each researcher and other prominent researchers (Fig.~3A-H). When counting collaborations, several repeated collaborations with the same alter prominent researcher are counted separately, so that one collaborator can give rise to several collaborations. Across all fields, we find that the average number of collaborations with other prominent researchers is always higher in Europe than in North America even though in fields with lower collaboration rates among prominent researchers the differences are not statistically significant (Fig.~3).
When all fields are combined (normalizing each field), the difference is significant at the 1\% level (Fig.~3I). Similarly, a significant  majority of researchers with above-median number of collaborations with other prominent researchers are based in Europe, whereas the majority of researchers with below-median number of collaborations with other prominent researchers are based in North America (Fig.~S10; Methods). Taken together with the fact that the total number of collaborators does not differ significantly between Europe and North America (Fig.~1 and Fig.~S1), these results indicate that researchers based in North America have a higher tendency to collaborate with non-prominent researchers, whereas in Europe the research elite in a specific field is more tightly knit.

Second, we measure, for each prominent researcher, the fraction of prominent researchers in their continent with which the researcher has collaborated. We call this the \textit{fraction of intracontinental collaborators} (Fig.~3J-Q); a value of 0.5 indicates that a prominent researcher has collaborated with half of the prominent researchers in their continent. If we pool all fields together, we find that the fraction of intracontinental collaborators normalized by field is significantly higher in Europe than in North America at $1\%$ level (Fig. 3R). For individual fields, we find that the mean fraction of intracontinental collaborators is always significantly higher in Europe than in North America and that prominent researchers in Europe have significantly above-median intracontinental collaborators for all fields (Fig. S10), except for  metabolomics (Methods).

\subsection*{Collaboration-dependent differences in impact}

If, as we have shown, collaboration patterns are different across continents, and if collaboration network structure affects research performance \cite{guimera05c, chan16,danchev19}, then we expect systematic collaboration-dependent differences in impact across continents. To investigate this question, we analyze the impact of publications of prominent researchers, both when they publish with and without other prominent researchers (Fig.~4 and Supplementary Fig.~S11). We quantify the value added by collaboration by means of the normalized logarithmic impact, which measures the (logarithmic) impact of papers relative to the (logarithmic) impact of papers written by single prominent researchers in the same year (Methods). We find that, in general, researchers in North America publish significantly more impactful papers than those in Europe when they publish without other prominent researchers in their field (in philosophy of science and cognitive psychology the differences are not significant). Since, as we have seen earlier, prominent researchers in Europe collaborate more with other prominent researchers (Fig.~4), this may provide a mechanism to compensate, by means of collaboration, for the lower impact of their work without other prominent researchers.

We also find that collaborating with other prominent researchers increases, by 15\% on average across all fields, the impact of publications  (differences not significant for Europe-based researchers in philosophy of science and genetics, and North America-based researchers in inequalities in health). The prominent researchers in Europe and North America who benefit the most, in terms of higher publication impact, by collaborating with other prominent researchers are those in the fields of network science, with an increase of 25\% and 33\%, and development economics, with 20\% and 30\%, respectively. 

However, previous results linking collaboration network structure to outcome quality \cite{guimera05c,hoekman12} have generally indicated that repetitive collaborations with the same researchers and largely closed collaboration networks (as those observed in Europe) result in lower reproducibility and impact. Given the observed differences in collaboration patterns between continents, we investigate in more depth the effect of repeated collaborations (collaboration number) on the value added by collaboration (Fig.~5 and Supplementary Figs.~S11-12). Specifically, we analyze the normalized logarithmic impact for the first two collaborations among each pair of prominent researchers, the third to fifth collaborations, and the sixth collaboration and higher. Although the numbers in each field are small, often leading to non-significant differences, when all fields are pooled together a clear and significant pattern emerges: the more times a collaboration is repeated, the lower the impact (with  collaborations between prominent researchers in North America always having higher normalized logarithmic impact). The first two collaborations among prominent researchers increase (on average) the impact with respect to papers with a single prominent researcher by 34\% for North America based and 23\% for Europe based researchers. For 3-5 (and 6 or more repeated collaborations) the increase in impact is lower: 29\% (21\%) and 22\% (12\%) for North America and Europe, respectively. Trends among Europeans and North Americans follow similar patterns  within all fields (with Europeans having overall slightly lower impact). Nonetheless the increase and subsequent decrease varies  across all fields. This suggests that the nature of the returns to repeated collaborations are also influenced by field-specific features and not just the overall research environment and the number of times researchers collaborate.

\section*{Discussion}

In studying complex systems like the scientific process or collaboration networks, we are often constrained in precisely measuring causal relations. Here, we surmised that collaboration networks and scientific impact differ systematically across regions, and we found that the empirical evidence indeed supports this hypothesis. This does not prove that the research environments in Europe and North America are directly responsible for the observed differences in collaboration structure (and, indirectly, impact); but considering that research environment is known to affect collaboration network structure in some cases \cite{srivastava11,uzzi13,bromham16,wu19}, we can conjecture about causal mechanisms that could potentially lead to some of the observed differences.

In Europe, relative scarcity of research funds, collaboration-by-design in framework programs, and the European Commission's funding schemes can in part account for the larger number of collaborations among Europeans and the formation of a close-knit network of prominent scientists \cite{hoenig17, olechnicka19}. This collaborative strategy has resulted in EU15 competing with the US as the world’s largest scientific producing block in the last decades \cite{king04,lepori19}, although East Asia is catching up quickly. Paradoxically, even if collaborative productivity increases, this does not necessarily imply greater impact since largely closed networks of prominent scientists in Europe could result in less original and impactful research \cite{guimera05c}. Indeed, as illustrated above, the US has systematically been found to be more impactful across scientific fields \cite{king04,tu19}.

Nonetheless, the observation that for Europe-based scientists there is an advantage to collaborating with prominent Europe-based scientists suggests that there might be other mechanisms at play that go beyond funding agency norms. Europeans for example have shorter average travel distances and  live in similar time zones, and North Americans are commonly viewed as slightly more competitive and self-confident in their work \cite{triandis01, nisbett03}. Citations, famously referred to by Merton as ``pellets of recognition,'' contribute to appointment and promotions decisions \cite{ioannidis19,baccini19}. A growing supply of scientists and a stagnant number of tenured positions in the last three decades has led to greater competition for good jobs among scientists in Europe, vis-à-vis North America \cite{maher16}.
%

In this context, for aspiring Europe-based scientists, co-authorship with prominent scientists might be a dominant and effective social mechanism of professional advancement to secure access to scarce tenured positions \cite{sekara18,calvo-armengol04}. On the other hand, in North America the existence of individual soft money for career promotion coupled with less secure and influential tenured positions \cite{parker10,beaver01} could lead to permeable networks which are more open to newcomers and with fewer incentives for social closure through collaborations with respect to Europe. In fact, in North America the competition for resources through soft-money positions, prestige of first and last-authorship (to which researchers often renounce in large collaborations), and individual rewards could be a deterrent for prominent scientists to engage in systematic collaborations with other prominent scientists \cite{beaver01}.
Social stratification is well known to play an important role in student acceptance and hiring inequalities \cite{massey06,bowles08,clauset15} and could also play an important role in shaping the collaborations that prominent scientists establish. Further studies based on our findings could examine which forms of social stratification result in differential access to networks of prominent scientists in North America and Europe.

More generally, statistical analysis of network structures  linked to impact of the scientific output can be limited in providing a precise causal mechanism given factors that are not easily measurable such as researchers' personality traits (such as being more competitive and self-confident) and individual motivations towards collaboration \cite{boyer17}, and the social norms that shape differences in scientific cultures across continents.
%
%
Nonetheless, our finding that research that involves several prominent researchers has larger impact, which however wanes in repeated collaborations, holds across regions and scientific fields can have important implications. On the one hand, the share of research funding allocated to teams (and to repeating teams) may need to be reassessed for existing funding schemes. On the other, early career researchers may need, given different hiring criteria across fields, to strike a balance between work they do by themselves and in collaboration.

\section*{Materials and methods}

\subsection*{Data acquisition}

For constructing the lists of prominent researchers, we used the following procedures. The list for social inequalities in health was previously collected by another research group in Ref.~\cite{bouchard15}. %
%
For the four more established fields in our analysis (genetics, development economics, philosophy of science, and cognitive psychology) we selected the 100 researchers with the highest h-index in their field using Google Scholar in January 2021. We then  confirmed our initial list using Scopus’ citation and h-index data. To ensure that all researchers commonly viewed as the most influential were included in the top 100 lists, we checked common rankings of the most influential researchers for each of these fields. We incorporated the few top researchers in these ranked lists, who were not already among the top researchers according to Google Scholar. As these four fields have a longer tradition than the other four fields, we only included researchers with publications between 1960 and 2021.

For the four younger fields, for those with well defined conferences and scientific societies (network science and metabolomics) we identified the main conferences (NetSci, NetSciX and CompleNet, for network science; and events of the Metabolomics Society for metabolomics) and societies (Network Science Society and Metabolomics Society), and considered all researchers who gave talks, are in scientific committees and scientific boards, and received awards in these venues. The authors in this list were identified in the Scopus database and ranked by their h-index. 
For the field of network ecology, we assembled the initial list by querying the Scopus database using a series of keywords (Ecologi* Network*, Food Web*, Environment* Network*, Trophic* Network*, Trophi* Web*) and focusing on top interdisciplinary and ecology journals. We then refined and ranked the list using the h-index, as before.

For all fields, we used Scopus database to extract all publications and bibliometric data for each author.
In all cases, we excluded the few researchers (a total of 6\%) not based in Europe or North America, or who did not collaborate with any other prominent researchers in the network. We also checked manually that all researchers in the network really have a significant body of work in the field, and excluded a few scientists that are prominent in other fields and have only made a small contribution to the field under consideration.

Note that we consider all the publications of authors that are prominent in each field, including publications in other fields. This is because we are interested in all the collaborations between these researchers. Additionally, we assign each prominent researcher to their main current affiliation, although some of them have developed parts of their careers in North America and Europe. 

The names of the prominent researchers are provided in Supplementary Figs.~S2-S9. Note that our criteria guarantee that all scientists in the network are prominent, although different criteria may result in somewhat different prominent researchers. Overall, all 100 researchers identified in each field are among the most highly cited and influential researchers with the highest h-index in their given field. We validated the data set by using an alternative method based on a Scopus search by keywords. The overlap with the network identified here, in network science for example, was 90\% and all results in the study remained the same.

\subsection*{Hierarchical stochastic block model for the analysis of network positions and roles}

Stochastic block models (SBM) are a class of  generative models for networks \cite{white76,nowicki01,guimera09}. In SBM, nodes are assumed to belong to groups, and node-to-node connectivity is defined by the group memberships alone. In particular, if nodes $i$ and $j$ belong to groups $g_i$ and $g_j$, respectively, then the probability that they are connected is given by a fixed number $p_{g_i g_j}$, which is identical for all other pairs in $g_i$ and $g_j$. The degree-corrected stochastic block model (SBM-DC) \cite{karrer11} is a variant of the SBM that allows for each node to have a different propensity to create links with others, thus allowing nodes in the same group to have broad degree (connectivity) distributions despite having the same connectivity patterns.

Because group memberships are typically unknown, it is necessary to infer them from the observed connections in a given network. The most plausible partition of the nodes into groups (the partition that maximizes the Bayesian posterior over partitions  \cite{guimera09}) is also the one with the minimum description length (MDL), that is, the one that most compresses the observed connections \cite{peixoto14}. To obtain the MDL partition, one needs to specify a prior distribution over partitions. A hierarchical prior, in which groups of nodes are assumed to be nested hierarchically, tends to yield shorter descriptions lengths than more uninformative priors. The SBM with such hierarchical priors is often referred to as the hierarchical SBM (hSBM) \cite{peixoto14}.

We obtain the MDL partition of the nodes in a collaboration network by using the \href{https://graph-tool.skewed.de/}{graph-tool Python module}. For each network, we fit the regular as well as the degree-corrected SBM, both with non-informative and hierarchical priors. In all cases (Supplementary Tables~S1,2), the degree-corrected hSBM yields the shortest description length, so all results reported in the manuscript correspond to this model. In Fig.~2, we use the groups at the second level of the hierarchy.

\subsection*{Group polarization and statistical significance of North America-Europe differences}

To quantify the affiliation imbalance of the groups identified by the hSBM, we defined group polarization $g_{p}$ as follows. For each researcher $i$ in a group of prominent researchers, we calculated the fraction of others in the group that belong to the same continent as $i$. Then, the mean group polarization $g_p$ is calculated as a mean over all researchers in all groups:
\begin{equation}
g_p =\frac{1}{N}\sum_{g = 1}^{G}\sum_{i=1}^{n_g} \frac{c_{ig} - 1}{n_g-1} 
\label{eq-gpolarity}
\end{equation}{}
where $N$ the number of researchers in the network, $G$ is the number of groups, ${c_{ig}}$ the number of researchers in group $g$ (other than $i$) belonging to the same continent than node $i$, and $n_{g}$ the total number of nodes in group $g$. Thus, the polarization of the network is $g_p=1$ if all groups comprise researchers only from North America or only from Europe, but no group contains researchers from both.

To assess the statistical significance of the group polarization of a given partition of the collaboration network into groups, we used resampling. In particular, we randomly reassigned the affiliations of all researchers in the network and calculated $g_p$ maintaining the same groups, and repeated this operation many times to obtain the null distribution of $g_p$ (Fig.~2C and Supplementary Figs.~S2-9C). 

To estimate the significance of the differences between North America and Europe we used, in all cases, resampling. In particular, for a given metric (number of authors per paper, log number of prominent researcher collaborations, fraction of intracontinental collaborators) we used the difference in the means as our statistic, and then obtained the expected null distribution of the statistic (and the p-value) by randomly reassigning affiliations to authors many times. Similarly, to establish the significance of the differences between fractions of above- and below-median researchers (for log number of prominent researcher collaborations and fraction of intracontinental collaborators) we calculated the actual value of such differences and compared them to the null expectation obtained by, again, reshuffling affiliations.

For the joint distribution of Fig.~1 we used the 2D Kolmogorov-Smirnov statistic \cite{press88} and, once more, we obtained the significance levels by repeatedly reshuffling the affiliations of each researcher.

\subsection*{Distribution of the logarithmic number of citations}

Fig.~4 shows the mean log number of citations for papers published by one or multiple prominent researchers, and by researchers in North America, Europe, or both. Supplementary Fig.~S11 shows the whole distribution of the logarithmic number of citations for the same papers. This distributions were compared using the Kolmogorov-Smirnov statistic and, as everywhere else, the significance of the observed differences was calculated by repeatedly reshuffling researchers' affiliations.

\subsection*{Normalized logarithmic impact}

To measure how research impact varies when prominent researchers collaborate (Figs.~4 and 5), we use normalized logarithmic impact as defined next. The normalized logarithmic impact $I_i$ of a paper $i$ is the logarithmic number of citations (plus 1) $\log (k_i + 1)$ of the paper divided divided by the mean of the logarithmic number of citations (plus 1) of papers with no prominent researcher collaboration in the same publication year
\begin{equation}
    I_i = \frac{\log (k_i + 1)}{\langle \log (k + 1)\rangle_{y_i}}\,.
\end{equation}
Here $\langle\dots\rangle_{y}$ is the mean over all papers published in year $y$ by single prominent researchers (and, possibly, other non-prominent researchers, but not multiple prominent researchers). Comparison to publications in the same year is necessary to avoid the artifact of later collaborations being less impactful just because they have had less time to accrue citations.

\bibliographystyle{Science.bst}
\bibliography{ref-collaborators.bib}

\section*{Acknowledgments}

We acknowledge Louise Boucchard for  sharing her complete list of authors with CM. This project has received funding from the Spanish Ministerio de Economía y Competitividad (PID2019-106811GB-C31) and from the Government of Catalonia (2017SGR-896).

\subsection*{Authors contributions}
CM, MS-P, RG formulated the overarching research goals and aims. LD and AK collected the data. LD developed the computer code for data analysis and visualization, and conducted the experiments. All authors analyzed and discussed the data and wrote the manuscript.

\subsection*{Competing interests}
The authors have no competing interests.

\subsubsection*{Data and materials availability}
All data are publicly available, and the lists of prominent researchers and their publications will be provided upon request.

\section*{Supplementary materials}
Figs. S1 to S12\\
Tables S1 and S2 \\

\clearpage

\section*{Figures}

\begin{figure}[h!]
\centerline{\includegraphics[width=1\textwidth]{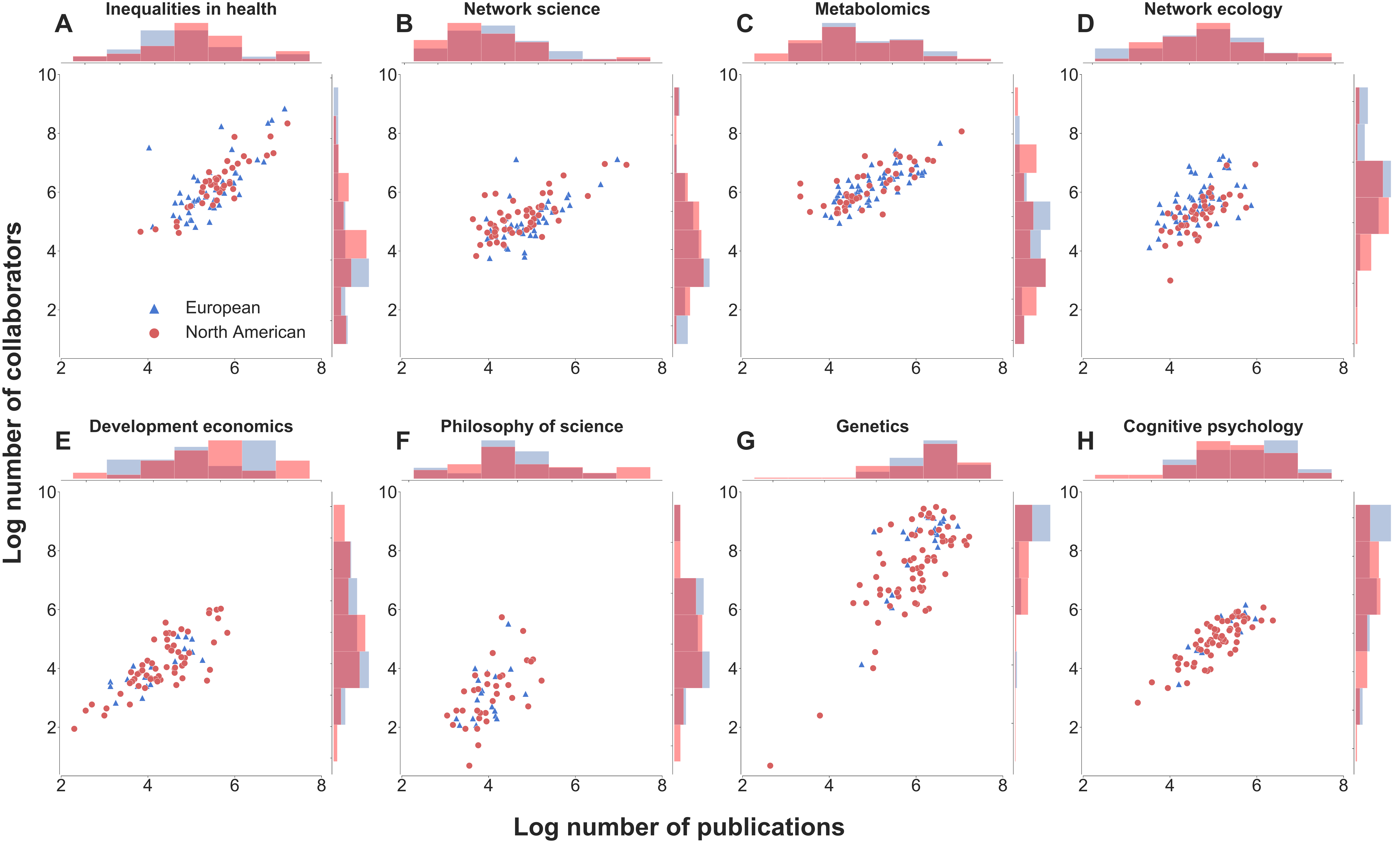}}
\renewcommand{\baselinestretch}{1.0}
\caption{\footnotesize
{\bf Total number of collaborators and publications for prominent researchers.}
{\bf (A) }-{\bf (H)} Logarithm of the total number of collaborators as a function of the logarithm of the total number of publications for each prominent researcher in:  {\bf (A)} inequalities in health, {\bf (B)} network science, {\bf (C)} metabolomics, , {\bf (D)} network ecology, {\bf (E)} development economics, {\bf (F)} philosphy of science, {\bf (G)} genetics and {\bf (H)} cognitive psycology. Red circles and blue triangles correspond to prominent researchers based in North-America and Europe, respectively. We test whether the points are distributed differently using the 2D Kolmogorov-Smirnov statistic \cite{press88}, and calculate the significance by resampling the researchers' affiliations. At the 5\% confidence level, we can only reject the null hypothesis (that both subsets are drawn from the same distribution) in the case of network ecology.
\label{fig-scatters}
}
\end{figure}
%

%
%
%
%

\clearpage

\begin{figure}[t!]
\includegraphics[width=\textwidth]{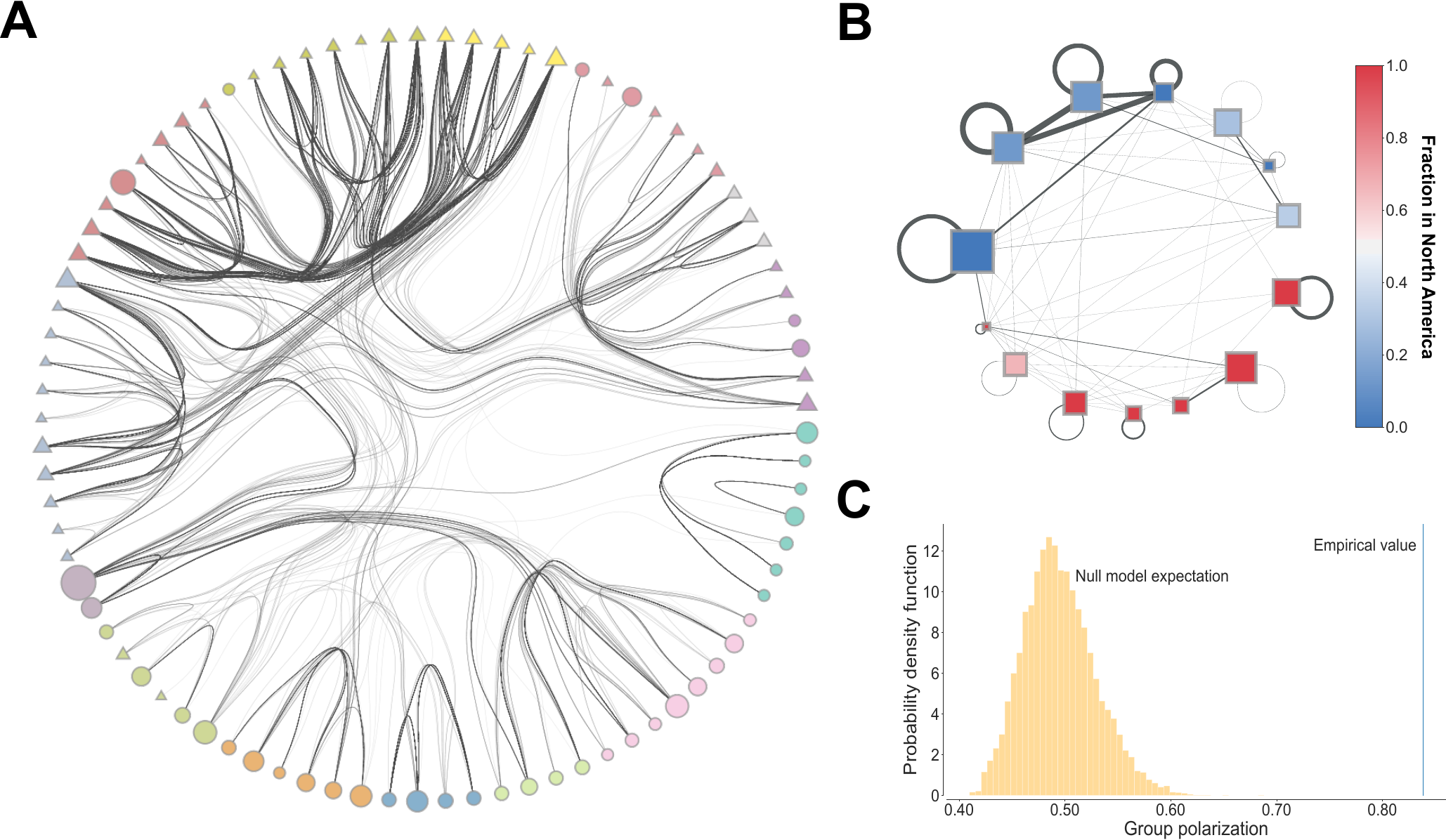}
\vspace{-7mm}
\renewcommand{\baselinestretch}{1.0}
\caption{\footnotesize
{\bf Stochastic block model and group polarization for the collaboration network in the field of inequalities in health.} (See Supplementary Figs.~S2-S8 for all other fields).
{\bf (A)} Collaboration network and best fit of the hierarchical stochastic block model (hSBM). Each node in the network represents a prominent researcher, and each edge represents a different collaboration (coauthored paper) between a pair of researchers. Prominent researchers  in North America and Europe are represented as circles and triangles, respectively. Different colors correspond to the groups identified by the hSBM, so that nodes with the same color have a similar collaboration pattern with other researchers and therefore fulfill a similar structural role in the collaboration network. Node size represents the betweenness centrality of the researcher in the network. We omit the names of the researchers so as not to distract from the patterns we aim to explore; for reproducibility, we provide the names of the prominent researchers in Extended Data Fig.~1, and in Supplementary Materials Fig.~S2-S8 for the other fields.
{\bf (B)} Block model of the collaboration network. Each node represents a group of researchers with similar collaboration patterns (that is, a different color in {\bf (A))}, with node size representing the number of researchers in the group. The width of the edges represents the number of collaborations between groups, and loops represent collaborations within each group. The color of each node indicates the fraction of researchers in the group that are based in North America, so that dark blue nodes represent groups with mostly Europe-based researchers, and red nodes represent groups with mostly North America-based researchers.
{\bf (C)} We define the polarization of a group as the number of same-continent researchers in the group over the random expectation for such number (Methods). The vertical line indicates the mean group polarization for the observed collaboration network. We randomize authors' affiliations and calculate the distribution of expected (null) polarization values. The empirical value is well above the null expectation, so that the group structure of the observed network is significantly polarized.
}
\label{fig-blocks}
\end{figure}

\clearpage

\begin{figure}[h!]
\centerline{\includegraphics[width=\textwidth]{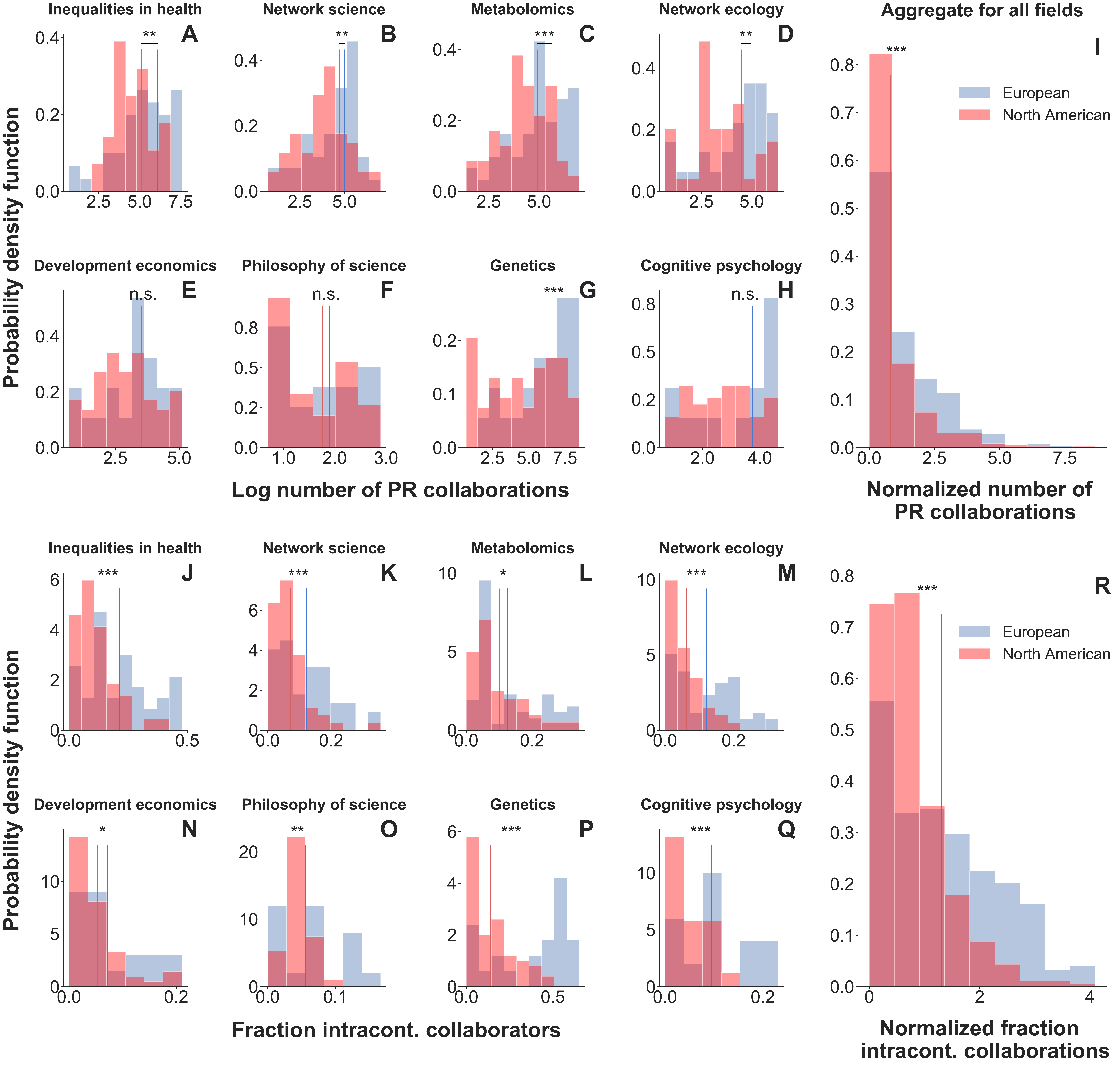}}
\renewcommand{\baselinestretch}{1.0}
\caption{\footnotesize
{\bf Differences in collaboration patterns between prominent researchers in North America and Europe.}
{\bf (A)}-{\bf (H)} Number of prominent researcher (PR) collaborations. We plot the distribution of the logarithm of number collaborations for prominent researchers in North America (red) and Europe (blue). The vertical lines indicate the mean log-number of collaborations for each subset. The significance of the difference between the European and North American means was established by resampling researcher affiliations (one sided test). {\bf (I)} Aggregated distribution for all fields. The log-number of PR collaborations are normalized by the mean in each field so as to make all fields comparable.
{\bf (J)}-{\bf (Q)} Fraction of intracontinental collaborators, defined as the fraction of prominent researchers in the same continent with which a prominent researcher collaborates. We plot the distribution of the fraction of intracontinental collaborators in North America (red) and Europe (blue). The vertical lines indicate the mean fraction of intracontinental collaborators for each subset. The significance of the difference between the European and North American means was established by reshuffling researcher affiliations (one sided test). {\bf (R)} Aggregated distribution for all fields. The fractions of intracontinental collaborations in each field are normalized by the mean of the field so as to make all fields comparable.
Stars indicate significant differences (***: 1\%, **: 5\%, *: 10\%, n.s.: not significant).
}
\end{figure}

\clearpage

\begin{figure}[h!]
\centerline{\includegraphics[width=1.\textwidth]{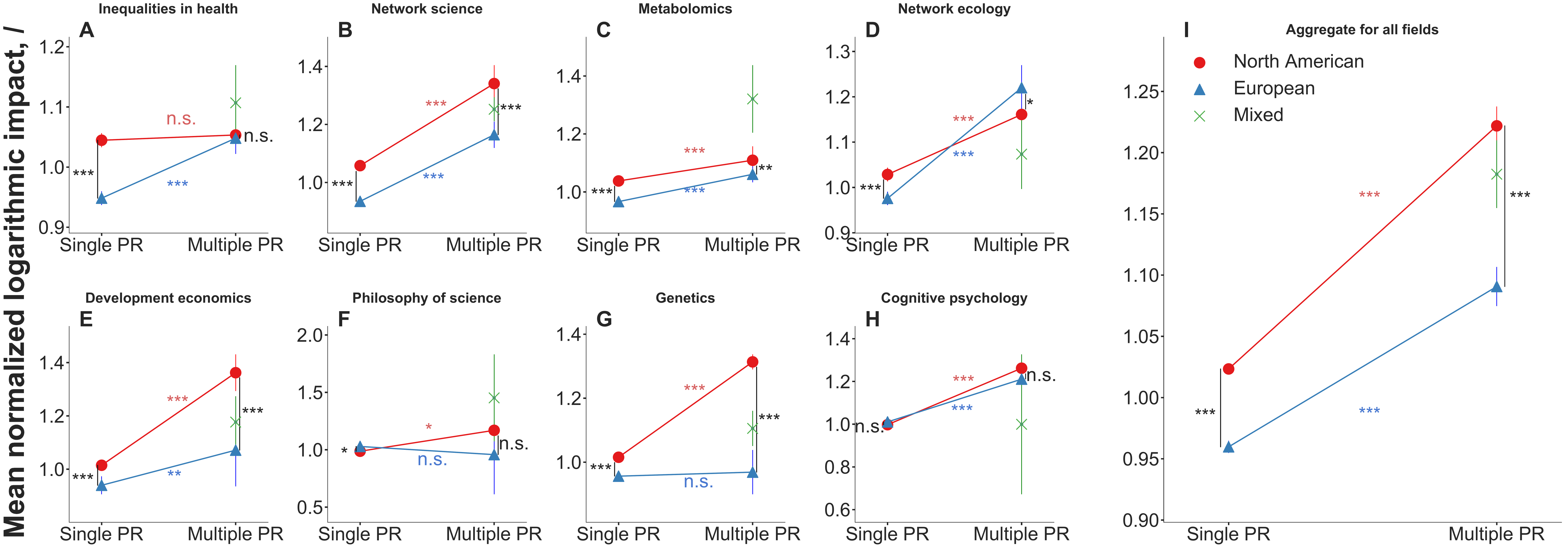}}
\renewcommand{\baselinestretch}{1.0}
\caption{\footnotesize
{\bf Impact difference under different collaborative strategies.}
Mean normalized logarithmic impact for articles authored by either a single prominent researcher (PR) or multiple PR in: {\bf (A)} inequalities in health, {\bf (B)} network science, {\bf (C)} metabolomics, {\bf (D)} network ecology, {\bf (E)} development economics, {\bf (F)} philosophy of science, {\bf (G)} genetics and {\bf (H)} cognitive psychology.
The normalized logarithmic impact $I_i$ of a paper $i$ is the logarithmic number of citations (plus 1) $\log(k_i+ 1)$ of the paper divided by the mean of the logarithmic number of citations (plus 1) of papers with no prominent researcher collaboration in the same publication year (Methods).
{\bf (I)} Aggregated normalized logarithmic impact for all fields.
Stars indicate significant differences (***: 1\%, **: 5\%, *: 10\%, n.s.: not significant). See Extended Data Fig.~5 for the whole distributions of the logarithmic number of citations.
}

\label{fig-impact}
\end{figure}

\clearpage

\begin{figure}[h!]
\centerline{\includegraphics[width=1.2\textwidth]{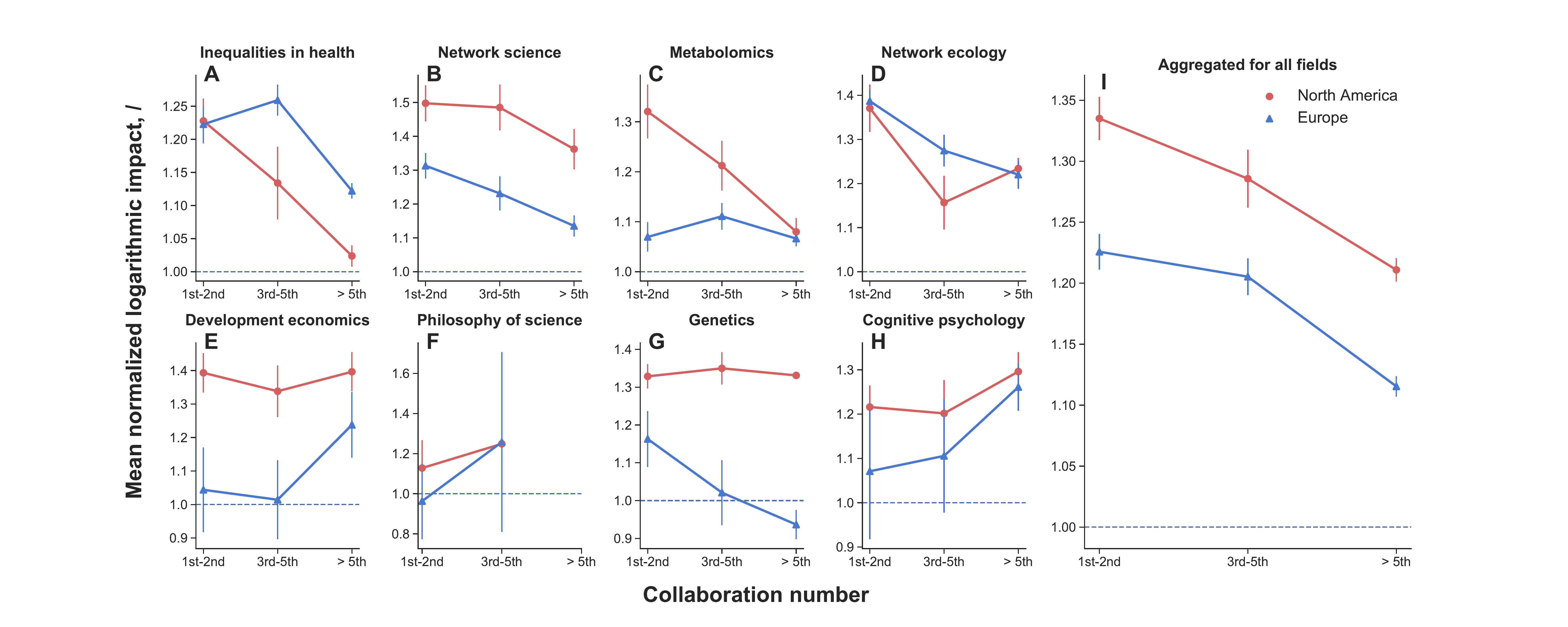}}
\renewcommand{\baselinestretch}{1.0}
\caption{\footnotesize
{\bf Evolution of impact with repeated collaborations.}
Mean normalized logarithmic impact of publications authored by a pair of prominent researchers (PR) as a function of the collaboration number (the number of times two prominent researchers have co-authored a paper: 1-2, 3-5, or $>$5; Methods).
The normalized logarithmic impact $I_i$ of a paper $i$ is the logarithmic number of citations (plus 1) $\log(k_i+ 1)$ of the paper divided by the mean of the logarithmic number of citations (plus 1) of papers with no prominent researcher collaboration in the same publication year (Methods).
{\bf (A)} inequalities in health, {\bf (B)} network science, {\bf (C)} network ecology, {\bf (D)} metabolomics, {\bf (E)} development economics, {\bf (F)} philosophy of science, {\bf (G)} genetics and {\bf (H)} cognitive psychology.
{\bf (I)} Aggregated normalized logarithmic impact for all fields, as a function of the number of collaboration.
}

\label{fig-repeated}
\end{figure}

\end{document}